\documentclass[structabstract]{aa}  

\usepackage{graphicx}
\usepackage{epsfig}
\usepackage{txfonts}
\usepackage{natbib}
\bibpunct{(}{)}{;}{a}{}{,} 

\def\apj{{\em ApJ}}
\def\apjs{{\em ApJS}}
\def\apjl{{\em ApJ}}
\def\aap{{\em A\&A}}
\def\aj{{\em AJ}}

\def\mnras{{\em MNRAS}}

\def\araa{Annual Review of Astron and Astrophys}
 
\def\aaps{{\em A\&AS}} 
\newcommand{\lunits}{\ensuremath{\mathrm{erg\,s^{-1}\,Hz^{-1}}}}
\newcommand{\lsun}{\,\mbox{\ensuremath{\mathrm{L_{\odot}}}}}       
\newcommand{\lir}{\ensuremath{L_{\mathrm{IR}}}}                    
\newcommand{\mum}{\,\mbox{\ensuremath{\mathrm{\mu m}}}}           
\newcommand{\snrate}{\ensuremath{\nu_{\mathrm{SN}}}}  
\newcommand{\mujyb}{\,\mbox{\ensuremath{\mathrm{\mu Jy/beam}}}}    
\newcommand{\mjyb}{\,\mbox{\ensuremath{\mathrm{mJy/beam}}}}        
\newcommand{\jyb}{\,\mbox{\ensuremath{\mathrm{Jy/beam}}}}          
\newcommand{\rlunits}{\ensuremath{\,\mathrm{erg\,s^{-1}}}}        
                    
\begin{document}
   \title{e-MERLIN and VLBI observations of the luminous infrared galaxy IC\,883: a nuclear starburst and an AGN candidate revealed}

   \author{C. Romero-Ca\~nizales \inst{1,2,}\thanks{\email{crroca@utu.fi}}
          \and
           M.\,A. P\'erez-Torres \inst{1}
          \and
          A. Alberdi \inst{1}
          \and
          M.\,K. Argo \inst{3,4}
          \and
          R.\,J. Beswick \inst{4}
          \and
          E. Kankare \inst{2}
          \and
          F. Batejat \inst{5}
          \and
          A. Efstathiou \inst{6}
          \and
          S. Mattila \inst{2}
          \and
          J.\,E. Conway \inst{5}
          \and
          S.\,T. Garrington \inst{4}
          \and
          T.\,W.\,B. Muxlow \inst{4}
          \and
          S.\,D. Ryder \inst{7}
          \and
          P. V\"ais\"anen \inst{8}
          }

    \institute{
             Instituto de Astrof\'{\i}sica de Andaluc\'{\i}a - CSIC, PO Box 3004, 18008 Granada,  Spain
         \and
             Tuorla Observatory, Department of Physics and Astronomy, University of Turku, V\"ais\"al\"antie 20, 
             FI-21500 Piikki\"o, Finland 
         \and
             Netherlands Institute for Radio Astronomy (ASTRON), Postbus 2, 7990 AA Dwingeloo, The Netherlands
         \and
             Jodrell Bank Centre for Astrophysics, Alan Turing Building, School of Physics and Astronomy, The 
             University of Manchester, Manchester M13 9PL
         \and
             Onsala Space Observatory, SE-439 92 Onsala, Sweden
         \and
             School of Sciences, European University Cyprus, Diogenes Street, Engomi, 1516 Nicosia, Cyprus
         \and
            Australian Astronomical Observatory, PO Box 296, Epping, NSW 1710, Australia            
         \and
            South African Astronomical Observatory, PO Box 9, Observatory 7935, South Africa
              }

\date{Received \today; accepted ?}

 
  \abstract
{The high star formation rates of luminous infrared galaxies (LIRGs) make them ideal places 
for core-collapse supernova (CCSN) searches. Massive star formation can often be found 
in coexistence with an active galactic nucleus (AGN), contributing jointly to the energy 
source of LIRGs. At radio frequencies, free from dust extinction, it is possible to detect 
compact components within the innermost LIRG nuclear regions, such as SNe and SN remnants, as 
well as AGN buried deep in the LIRG nuclei.}
{Our study of the LIRG IC\,883 aims at: (i) investigating the parsec-scale radio structure of the 
(circum-)nuclear regions of IC\,883; (ii) detecting at radio frequencies the two recently reported 
circumnuclear SNe 2010cu and 2011hi, which were discovered by near-IR (NIR) adaptive optics observations 
of IC\,883; and (iii) further investigating the nature of SN\,2011hi at NIR wavelengths.}
{We used the e-EVN at 5\,GHz, and e-MERLIN at 6.9\,GHz, to observe contemporaneously the LIRG 
IC\,883 at high angular-resolution (from tens to hundreds of milliarcsec) and with high sensitivity 
($<70\,\mu$Jy), complemented by archival VLBI data at 5 and 8.4\,GHz. We also used the Gemini-North 
telescope to obtain late-time \textit{JHK} photometry for SN\,2011hi.}
{The circumnuclear regions traced by e-MERLIN at 6.9\,GHz have an extension of $\sim$1\,kpc, at a 
position angle of 130\degr{}, and show a striking double-sided structure, which very likely corresponds 
to a warped rotating ring, in agreement with previous studies. Our e-EVN observations at 5\,GHz
and complementary archival VLBI data at 5\,GHz and 8.4\,GHz, reveal the presence of various milliarcsec 
compact components in the nucleus of IC\,883. A single compact source, an AGN candidate, dominates the 
emission at both nuclear and circumnuclear scales, as imaged with the e-EVN and e-MERLIN, respectively. 
The other milliarcsec components are very suggestive of ongoing nuclear CCSN activity. Our e-EVN 
observations also resulted in upper limits to the radio luminosity of the two SNe in IC\,883 recently 
discovered at NIR wavelengths. We refine the classification of SN\,2011hi as a Type IIP SN according to 
our latest Gemini-North epoch from 2012, in agreement with a low-luminosity radio SN nature. We estimate 
a CCSN rate lower limit of $1.1_{-0.6}^{+1.3}$\,yr$^{-1}$ for the entire galaxy, based on three nuclear 
radio SNe and the circumnuclear SNe 2010cu and 2011hi.}
{}


\titlerunning{IC\,883 revealed by e-MERLIN and VLBI observations}
\authorrunning{C. Romero-Ca\~nizales et al.}
   \maketitle
%

\section{Introduction}\label{sec:intro}

The excess of radio continuum and infrared (IR) emission in luminous IR galaxies (LIRGs) 
is associated with vigorous formation of massive stars ($M\ga8$\,M$_{\sun}$) and/or an active 
galactic nucleus (AGN) \citep[see e.g.,][]{sanders}. Core-collapse supernovae (CCSNe) serve as 
probes of the star formation rate (SFR) of massive stars, and therefore their detection in LIRGs 
is a pertinent task. 

IC\,883 is a LIRG with IR luminosity ($\lir=L[8$--$1000\,\mum]$) $\sim$$4.7\times10^{11}$\,\lsun{} 
at a distance of 100\,Mpc (1\,mas $\approx0.48$\,pc), according to \citet{sanders03}. This LIRG
is probably the result of the merger of two disk galaxies \citep{smithda95}, as suggested by its 
associated tidal tails: a diffuse one approximately perpendicular to the region where gas and dust
preferentially concentrate (i.e., the disk) as seen in optical and near-IR (NIR) studies 
\citep[e.g.,][]{smithda95,scoville00,modica11}, and a narrow one along the disk itself \citep{keel85}. 

Taking \lir{} as a measure of the rate at which massive stars are formed, 
results in a corresponding CCSN rate, \snrate{}, of $\sim$1.3\,yr$^{-1}$ for IC\,883 \citep[assuming 
the empirical relation obtained by][]{seppo01}. Optical spectroscopic studies by \citet{vei95} have led 
to the classification of the IC\,883 nucleus as a low-ionisation narrow emission-line region. However, 
\citet{yuan10} have recently reclassified this galaxy as a composite starburst-AGN, based on the revised 
optical diagnostic diagrams by \citet{kewley06}. The multi-wavelength study on IC\,883 by \citet{modica11} 
also supports a composite nature. Consequently it is expected that an AGN is contributing to the total 
\lir{} in IC\,883, and thus the \snrate{} should be lower than 1.3\,yr$^{-1}$. Indeed, at radio 
frequencies, \citet{sll98} and \citet{parra10} have found evidence of compact AGN activity through VLBI 
studies of IC\,883.

IC\,883 is part of our program ``An ALTAIR study of Supernovae in Luminous Infrared Galaxies'', 
using Gemini-North with its Laser Guide Star adaptive optics (AO) system \citep[e.g.,][]{erkki08}. During 
the four year survey, our program has yielded the discovery of two CCSNe in IC\,883 within a 12-month period: 
SN\,2010cu \citep[24 Feb. 2010,][]{sn10cu} and SN\,2011hi \citep[11 Feb. 2011,][]{sn11hi}. \citet{erkki12}
concluded that either a Type IIP or a IIn/L SN provided the best fits to the NIR light curves of SNe 2010cu 
and 2011hi.

In this paper we report our radio observations towards the nuclear and circumnuclear regions of IC\,883, 
as well as the most recent NIR epoch obtained with Gemini-North on January 31 2012, not included in 
\citet{erkki12}. In Section \ref{sec:ic883_obs} we give details of our radio and NIR observations, Section 
\ref{sec:ic883_results} contains our results and discussion, and Section \ref{sec:sum} contains the summary of 
our study.

\section{Observations}\label{sec:ic883_obs}

IC\,883 was observed at radio frequencies with the electronic Multi-Element Remotely Linked Interferometer 
Network (e-MERLIN) and with the electronic European very long baseline interferometry (VLBI) Network (e-EVN\footnote{The 
development of e-VLBI within the EVN has been made possible via EXPReS project funded by the EC FP6 IST Integrated 
Infrastructure Initiative contract \# 026642 - with a goal to achieve 1 Gbps e-VLBI real-time data transfer and correlation.}),
and at NIR wavelengths with the Gemini-North telescope. Our observations aimed at imaging the parsec-scale radio 
structure of the (circum-)nuclear regions of IC\,883, and detecting SNe 2010cu and 2011hi at radio frequencies, as well as to 
further investigating the nature of SN\,2011hi at NIR wavelengths. To complement our radio observations, we have also analysed 
publicly available archival VLBI data.

In Table \ref{tab:observ} we give some details on the observational parameters of the different radio 
observations included in our study: e-MERLIN, e-EVN and additional VLBA and EVN observations.

\begin{table*}
\caption{Observational parameters of the radio observations.}
\label{tab:observ}      
\centering          
\begin{tabular}{c c r c c c c c r c}    
\hline\hline  
 \multicolumn{1}{c}{Epoch} & \multicolumn{1}{c}{Project} & \multicolumn{1}{c}{Array} & \multicolumn{1}{c}{Observing} & 
   \multicolumn{1}{c}{$\nu$} & \multicolumn{1}{c}{Phase} & \multicolumn{1}{c}{$P_{\nu}$}  & \multicolumn{1}{c}{Weighting}    &
    \multicolumn{1}{c}{Convolving beam} & \multicolumn{1}{c}{rms} \\
\multicolumn{1}{c}{label}  &  &   & \multicolumn{1}{c}{date}  & \multicolumn{1}{c}{(GHz)} & \multicolumn{1}{c}{calibrator}  
  & \multicolumn{1}{c}{(\jyb{})} & ~ & \multicolumn{1}{c}{(mas)} & \multicolumn{1}{c}{(\mujyb{})}\\
\hline
\multicolumn{1}{c}{(1)} & \multicolumn{1}{c}{(2)} & \multicolumn{1}{c}{(3)} & \multicolumn{1}{c}{(4)} & \multicolumn{1}{c}{(5)} & 
\multicolumn{1}{c}{(6)} & \multicolumn{1}{c}{(7)} & \multicolumn{1}{c}{(8)} & \multicolumn{1}{c}{(9)} & \multicolumn{1}{c}{(10)} \\
\hline
EM & - & e-MERLIN & 2011-03-19 & 6.9 & J1324$+$3622  & 0.07 & N,1  & 165.23$\times$88.35 at $11.6\degr{}$    &  44.0 \\
V1 & BN026 & VLBA & 2004-08-13 & 5.0 & J1317$+$3425  & 0.25 & N,0  & 3.71$\times$1.47 at ~ $-0.1\degr{}$     &  84.7 \\
V2 & BN027 & VLBA & 2004-09-20 & 5.0 & J1310$+$3220  & 1.41 & N,1  & 3.48$\times$1.65 at ~ $17.6\degr{}$     & 119.0 \\
V3 & BN027 & VLBA & 2005-07-11 & 5.0 & J1310$+$3220  & 1.19 & N,2  & 3.43$\times$1.78 at ~ $11.5\degr{}$     & 107.0 \\
V4 & EP055 & EVN  & 2006-06-15 & 5.0 & J2333$+$3901  & 0.81 & U,1  &  7.99$\times$6.00 at $-81.3\degr{}$     & 162.0 \\
V5 & RR006 & e-EVN& 2011-03-23 & 5.0 & J1317$+$3425  & 0.35 & U,1  & 9.20$\times$6.36 at $-76.1\degr{}$      &  66.2 \\
V6 & BC196 & VLBA & 2011-05-15 & 8.4 & J2330$+$3348  & 0.33 & N,0  & 2.41$\times$1.17 at ~ $-8.3\degr{}$     & 174.0 \\
\hline              
\end{tabular}
\tablefoot{Phase reference sources used in each observation are listed along with their associated peak intensities 
(columns 6 and 7). For the different IC\,883 maps, column 8 lists the weighting parameters used in the imaging process: 
N$=$natural, U$=$uniform, and the number corresponds to the value of the ROBUST parameter). Column 9 lists the resolution, 
and column 10 the attained rms noise.}
\end{table*}

\subsection{e-MERLIN observations and data reduction}\label{sec:ic883mer_obs}

We carried out e-MERLIN observations of IC\,883 at 6.9\,GHz (median central frequency) on 19 March 2011. These 
were Director's Discretionary Time (DDT) observations within the commissioning phase of e-MERLIN, which included 
the following 25\,m diameter antennas: Mark II, Defford, Knockin, Darnhall and Pickmere. The observations lasted 
$\sim24$\,hr, from which approximately 20 were spent on target and 10\,hr were ultimately usable after 
editing. Four sub-bands (512 channels each) with dual polarisation were used, accounting for a total 
bandwidth of 512\,MHz.

We analysed the data within the NRAO Astronomical Image Processing System ({\sc aips}). 3C286 set the absolute 
flux density scale following an iterative process. The flux estimated for 3C286 in the shortest baseline (Mark 
II-Pickmere), at the centre of the different sub-bands, was on average 5.56\,Jy. We then calibrated the amplitude of 
DA\,193, which is a bright and unresolved source as seen by all e-MERLIN baselines, resulting in an average flux density
of 3.72\,Jy. Finally, we used DA\,193 to set the flux density of the phase reference source, J1324$+$3622, 
which resulted in an average flux density of 68.25\,mJy. We performed a series of phase-only self-calibration 
iterations of the phase reference source, before phase-calibrating the target source. To account 
for the correlation offset from the position of the strongest source in the field, we used the task 
UVSUB in {\sc aips}. To improve the sensitivity, we performed phase-only 
self-calibration iterations on the target source. We achieved a thermal rms noise in the IC\,883 map of 
44\,\mujyb{}, for a beam size of 165$\times$88\,mas at $-11.6\degr{}$ (see left panel in Figure 
\ref{fig:ic883zoom}), obtained with natural weighting and ROBUST=1 within {\sc aips}.

\subsection{e-EVN observations and data reduction}\label{sec:ic883evn_obs}

We observed IC\,883 on 23 March 2011 (ToO project: RR006, P.I.: Romero-Ca\~nizales) at 
5\,GHz, with the e-EVN, which included the following antennas (diameter, location): Effelsberg (100\,m, Germany), 
Mark II (25\,m, UK), Medicina (32\,m, Italy), Onsala (25\,m, Sweden), Torun (32\,m, Poland), Westerbork array 
(14$\times$25\,m, NL) and Yebes (40\,m, Spain). 

RR006 was a 2\,hr experiment ($\sim$1.3\,hr, total time on source), recorded at 1024\,Mbps 
using eight sub-bands, each of 16\,MHz and dual polarisation. The data were correlated at the EVN 
MkIV Data Processor at JIVE with an averaging time of 2\,s. The point-like source J1159$+$2914 
(2.45\,Jy at 5\,GHz) was used as a fringe finder and bandpass calibrator. J1317$+$3425 (0.35\,Jy at 
5\,GHz), at $\sim$0.7\degr{} angular distance from IC\,883, served as phase reference source. 
3.4\,min scans on IC\,883 were alternated with 1.2\,min scans on J1317$+$3425. The data were analysed 
with {\sc aips}. We also used the Caltech program {\sc DIFMAP} \citep{difmap} to image the calibrators and 
to assess the performance of each antenna. 

In an attempt to detect a radio counterpart for SN\,2011hi, its coordinates derived from the NIR images 
were used for pointing and correlation. A local rms of $\sim$30\,\mujyb{} was achieved at the pointing centre. 
Since no radio source was detected above $3\sigma$ at that position, we corrected the visibility data with 
the task UVSUB in {\sc aips}, as we did for the e-MERLIN data, to account for the correlation offset from the position 
of the strongest source in the field, i.e., to the radio nucleus of IC\,883. The shift of $\sim$0.8\,arcsec is 
well within the primary beam of each antenna  in the used array. 

For the imaging process within {\sc aips} we used different weighting schemes in order to test the reliability 
of our results. Due to the scarce uv-coverage, the use of natural weighting tends to give rise to secondary 
side lobes at a 60 per cent level of the main lobe, which can produce artifacts in the reconstruction imaging 
process that can be confused with putative sources. Our final image was made using uniform weighting and 
ROBUST$=$1. We achieved a final thermal rms in the map of 66.2\,\mujyb{}, for a beam size of 9.20$\times$6.36\,mas 
at $-76.1$\degr{} (see Figure \ref{fig:ic883zoom}, right panel).

\subsection{Archival VLBI data}

We have analysed Very Long Baseline Array (VLBA) data from projects BN026 and BN027 at 5\,GHz, and BC196 at 
8.4\,GHz. The VLBA observations under program BC196 were processed with the DiFX VLBI correlator \citep{difx}.
We also report EVN observations at 5\,GHz made on June 15th 2006 (project: EP055, P.I.: R.\,Parra); see Table
\ref{tab:observ}.

In project EP055, IC\,883  was observed in four scans spread over 24\,hr to optimize the uv-coverage for a total 
observation time on source of $\sim$1.3\,hr. The array used comprised 10 EVN stations: Lovell (76\,m, UK), Cambridge 
(32\,m, UK), Westerbork array (14$\times$25\,m, NL), Effelsberg (100\,m, Germany), Medicina (32\,m, Italy), Noto 
(32\,m, Italy), Onsala (25\,m, Sweden), Torun (32\,m, Poland), Urumqi (25\,m, China) and Shanghai (25\,m, China). 
The observations were performed in a 1024\,Mbps dual polarization mode. The data were correlated at JIVE in the 
Netherlands with an averaging time of 2\,s.

All the additional VLBI data were edited and reduced using standard procedures within {\sc aips}.
The phase reference calibrator used in project BN027 is slightly resolved at VLBA scales, and we have thus 
corrected the fringe solutions from its structure. We did not apply any self-calibration to obtain the final images 
of IC\,883. In the case of project EP055, the data from the two Chinese antennas (Ur and Sh) as well as from Cm could
not be properly calibrated, so we omitted these antennas in the imaging process.

As in the case of our e-EVN epoch, we have tested the reliability of the sources in the different VLBI
maps, by using different weighting schemes. We used the {\sc aips} task UVSUB to apply shifts in $\alpha$ and $\delta$ to 
the visibility data on all the additional archival VLBI epochs. This allowed us to have all the maps centred at the 
same position ($\alpha(\mathrm{J}2000) = 13^h20^m35\fs3184$, $\delta(\mathrm{J}2000) = 34\degr08\arcmin22\farcs352$). 
Since the resulting images have a different convolving beam and were thus mapped with different pixel sizes, we 
changed their geometry to match that of our e-EVN epoch (pixel size of 0.5\,mas), by means of the task OHGEO 
within {\sc aips}.

\subsection{Gemini-North observations}

Our Gemini-North observations of IC\,883 were conducted with the ALTAIR Laser Guide Star AO system on the 
Near-InfraRed Imager (NIRI). The LIRG was observed in 11 separate epochs between 2008 April 15.5 UT and 2012 January 
31.6 UT, covering a period of 3.8\,yr. Here we report the results from the most recent epoch of \textit{JHK} imaging 
obtained in 2012 under program GN-2012A-Q-56 (P.I.: S.\,Ryder); the previous epochs are reported in \citet{erkki12}. 
For details on the data reduction, calibration, image subtraction process and photometry of the ALTAIR/NIRI data on 
SN\,2010cu and SN\,2011hi we refer to \citet{erkki12}.

Image subtraction based on the ISIS 2.2 \citep{alard,alard00} package was used to compare the 2012 \textit{JHK} 
images of IC\,883 to the reference data obtained on 2010 May 4-5 UT. This revealed SN\,2011hi to have faded below the 
detection limit of the Gemini images. Similar to deriving the \textit{J}-band upper limit for SN\,2010cu in \citet{erkki12} 
we used the QUBA package \citep[see][]{valenti11} running standard {\sc iraf} tasks, to measure the faintest detectable 
sources in the 2010 reference image with poorer quality AO correction compared to the latest 2012 epoch. Taking into account 
the noise increasing by a factor of $\sqrt2$ due to the image subtraction process, conservative 5$\sigma$ upper limits of 
roughly $m_{J}>19.7$\,mag, $m_{H}>19.9$\,mag, and $m_{K}>19.9$\,mag were derived. 

\section{Results and discussion}\label{sec:ic883_results}

\begin{figure*}
   \centering
\includegraphics[bb= 6 3 778 411,clip,scale=0.65]{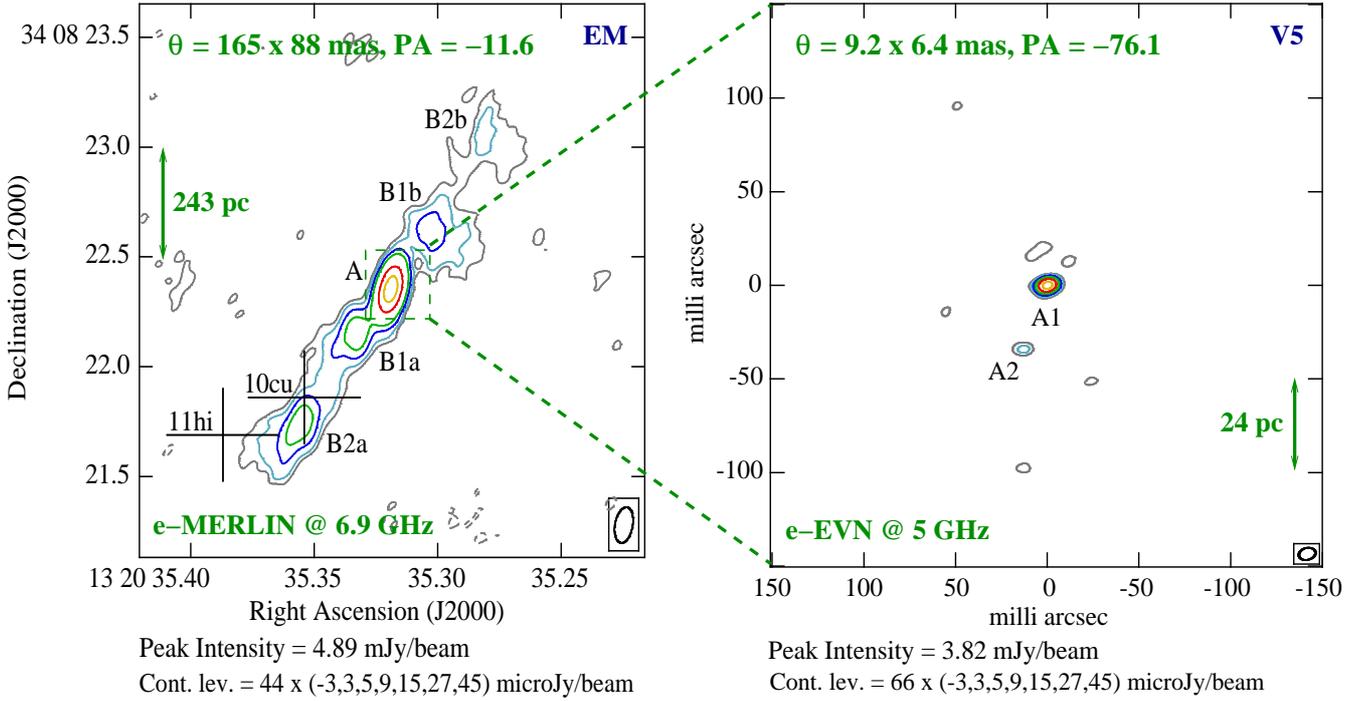}
   \caption{IC\,883 contour images at a median central frequency of 6.9\,GHz obtained with e-MERLIN (left), and at 5\,GHz 
            with the e-EVN (right), in observations carried out in March 2011. The rms noise in the two images is 44 and 
            66\,\mujyb{}, respectively, and the convolving beam in each case is indicated the upper left corner of the maps. 
            Dashed contours represent $-3\sigma$ levels. Note that the brightest component at e-MERLIN scales is dominated by 
            a compact source seen at mas scales with the e-EVN. The crosses in the left panel indicate the coordinates and 
            positional errors of SNe 2010cu and 2011hi, according to the values reported in \citet{erkki12}.}\label{fig:ic883zoom}%
\end{figure*}

\begin{figure*}[ht]
   \centering
\includegraphics[scale=0.58]{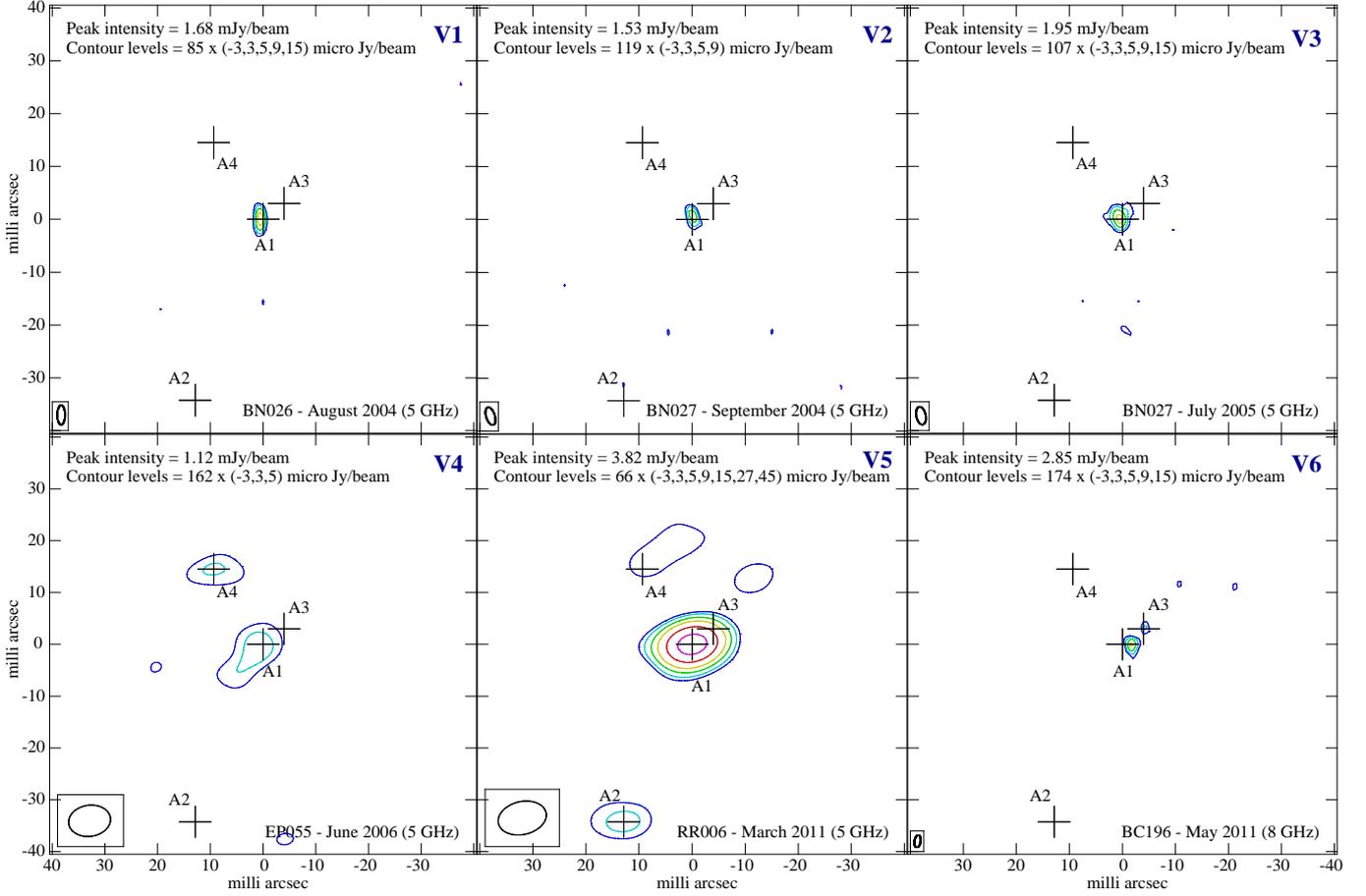}
   \caption{IC\,883 contour maps of the VLBI observations listed in Table \ref{tab:observ}. All the maps are centred on
  $\alpha(\mathrm{J}2000) = 13^h20^m35\fs3184$, $\delta(\mathrm{J}2000) = 34\degr08\arcmin22\farcs352$. The crosses indicate 
  the positions from the detection epoch of the different 5$\sigma$ sources reported in Table \ref{tab:ic883_radsou}: A1 and 
  A2 from experiment V5; A3 from experiment V6; and A4 from experiment V4.}\label{fig:allvlbi}
\end{figure*}

The images of the nuclear and circumnuclear regions of IC\,883 as observed with the e-EVN and e-MERLIN are
shown in Figure \ref{fig:ic883zoom}. In Figure \ref{fig:allvlbi} we show the images resulting from a total of six
VLBI epochs, covering a period of seven years. In Table \ref{tab:ic883_radsou} we report the estimated parameters 
for the $>5\sigma$ sources detected in all the images.

Our new Gemini-North epoch has yielded new \textit{JHK} magnitude limits for SN\,2011hi. In Figure \ref{fig:gemini}
we show the SN\,2011hi template light curve fits from \citet{erkki12}, including the new limits obtained from our 2012 
observations for comparison.

We note that the discovery of two SNe (2010cu and 2011hi) in a 3.8\,yr period results in a 
$\snrate{}\approx 0.5_{-0.3}^{+0.7}$\,yr$^{-1}$ \citep[using the upper and lower Poisson $1\sigma$ uncertainties 
for two events given by][]{gehrels}, which agrees within the uncertainties with the IR luminosity-based CCSN rate of 
the galaxy. However, this estimate does not include any correction factors nor control time considerations,
and should thus be regarded as a lower limit.

\begin{table*}
\caption{Estimated parameters of the detected radio emitting sources in the (circum-)nuclear regions of IC\,883.}
\label{tab:ic883_radsou}      
\centering          
\begin{tabular}{c c c c c c c c c}   
\hline\hline     
Source  & Epoch & $\Delta\alpha^s$(J2000) & $\Delta\delta^{\prime\prime}$(J2000) & $P_{\nu}$ & $S_{\nu}$ & $L_{\nu}$  &  
 $\Theta_{\mathrm{M}} \times \Theta_{\mathrm{m}}$ & log\,$T_{\mathrm{B}}$ \\
name    & label &  &  & (\mjyb{})  & (mJy)  & (10$^{27}$\lunits{}) &(mas$^2$)  & (K)  \\
\hline
(1) & (2) & (3) & (4) & (5) & (6) & (7) & (8) & (9)\\
\hline
A    & EM & 35.319 ($<$1) & 22.35 (1)  & 4.89 $\pm$ 0.25 & 6.07 $\pm$ 0.31 & 72.65 $\pm$ 3.67 & 104.8 $\times$ 30.4 & 4.15 $\pm$ 0.02  \\
B1a  & EM & 35.332 (1)    & 22.15 (3)  & 1.04 $\pm$ 0.07 & 2.07 $\pm$ 0.11 & 24.75 $\pm$ 1.34 & 138.1 $\times$ 97.1 & 3.45 $\pm$ 0.02  \\
B1b  & EM & 35.304 (1)    & 22.63 (6)  & 0.59 $\pm$ 0.05 & 1.75 $\pm$ 0.10 & 20.97 $\pm$ 1.17 & 191.3 $\times$ 141.6& 3.09 $\pm$ 0.02  \\
B2a  & EM & 35.356 (1)    & 21.74 (3)  & 1.03 $\pm$ 0.07 & 2.54 $\pm$ 0.13 & 30.34 $\pm$ 1.61 & 224.0 $\times$ 90.5 & 3.11 $\pm$ 0.02  \\
B2b  & EM & 35.281 (2)    & 23.05 (11) & 0.31 $\pm$ 0.05 & 0.97 $\pm$ 0.07 & 11.55 $\pm$ 0.78 & 295.9 $\times$ 109.8& 2.45 $\pm$ 0.03  \\
\hline 
A1 & V1 & 31.3180  ($\sim$0.0)& 22.352 (0.1)      & 1.68 $\pm$ 0.12 & 1.94 $\pm$ 0.13 & 23.19 $\pm$ 1.54 &  1.2 $\times$ 0.6      & 7.81 $\pm$ 0.03  \\
   & V2 & 31.3179 (0.1)       & 22.352 ($\sim$0.0)& 1.53 $\pm$ 0.14 & 1.53 $\pm$ 0.14 & 18.33 $\pm$ 1.69 &  0.6 $\times$ $\cdots$ & 6.79 $\pm$ 0.04  \\
   & V3 & 35.3181 (0.1)       & 22.349 (0.1)      & 1.95 $\pm$ 0.15 & 2.65 $\pm$ 0.17 & 31.73 $\pm$ 2.04 &  1.8 $\times$ 0.3      & 7.59 $\pm$ 0.03  \\
   & V4 & 35.3180 (0.6)       & 22.351 ($\sim$0.0)& 1.12 $\pm$ 0.17 & 1.21 $\pm$ 0.17 & 14.49 $\pm$ 2.07 & 13.0 $\times$ $\cdots$ & 5.97 $\pm$ 0.06  \\
   & V5 & 35.3179 (0.1)       & 22.352 ($\sim$0.0)& 3.82 $\pm$ 0.20 & 3.82 $\pm$ 0.20 & 45.67 $\pm$ 2.42 &  0.8 $\times$ $\cdots$ & 6.35 $\pm$ 0.02  \\
   & V6 & 35.3178 ($\sim$0.0) & 22.351 (0.1)      & 2.85 $\pm$ 0.22 & 4.38 $\pm$ 0.28 & 52.34 $\pm$ 3.34 &  1.5 $\times$ 0.4      & 7.55 $\pm$ 0.03  \\
\hline
A2 & V5 & 35.3190 (0.7)       & 22.317 (0.2)      & 0.42 $\pm$ 0.07 & 0.56 $\pm$ 0.07 &  6.66 $\pm$ 0.86 &  7.0 $\times$ 2.0      & 5.75 $\pm$ 0.06  \\
\hline
A3 & V6 & 35.3176 ($\sim$0.0) & 22.354 (0.2)      & 0.97 $\pm$ 0.18 & 1.55 $\pm$ 0.19 & 18.54 $\pm$ 2.28 &  1.4 $\times$ 1.1      & 7.15 $\pm$ 0.05  \\
\hline
A4 & V4 & 35.3181 (0.7)       & 22.371 (0.1)      & 0.89 $\pm$ 0.17 & 1.27 $\pm$ 0.17 & 15.25 $\pm$ 2.08 &  8.1 $\times$ 0.6      & 5.98 $\pm$ 0.06  \\
   & V5 & 35.3187 (0.9)       & 22.366 (0.2)      & 0.33 $\pm$ 0.07 & 0.67 $\pm$ 0.07 &  8.06 $\pm$ 0.89 & 14.0 $\times$ 1.4      & 5.23 $\pm$ 0.05  \\
\hline              
\end{tabular}
\tablefoot{
{\it Columns}: (1) Source names corresponding to those in Figures \ref{fig:ic883zoom} and \ref{fig:allvlbi}. (2) Epoch label 
corresponding to those specified in Table \ref{tab:observ}. Note that epoch EM and V5 are quasi-simultaneous. (3--4) 
Coordinates given with respect to $\alpha(J2000) = 13^h20^m00\fs0$ and $\delta(J2000) = 34\degr08\arcmin00\farcs0$. Position 
uncertainties in mas, within parentheses, are given by FWHM$/(2\times SNR)$, where $SNR$ is the signal to noise ratio, and FWHM 
was taken as the projection of the beam major axis on both $\alpha$ and $\delta$ axes. (5) Peak intensity. We estimated the 
uncertainties by adding in quadrature the rms noise in the map (column 10 in Table \ref{tab:observ}) plus a 5\% uncertainty 
in the point source flux density calibration. (6) Flux density. (7) Monochromatic luminosity at the observed frequency (column 
5 in Table \ref{tab:observ}). (8) Deconvolved major and minor axes, obtained by fitting a Gaussian to each source. (9) Brightness
temperature. When no deconvolved sizes could be obtained, we used instead the solid angle subtended by the synthesised beam for 
calculating lower limits of $T_{\mathrm{B}}$, i.e., 
$\Omega_{\mathrm{s}}=\pi(4\mathrm{log}2)^{-1}\left(\mathrm{FWHM}_{\mathrm{M}}\times\mathrm{FWHM}_{\mathrm{m}}\right)$, 
with $\mathrm{FWHM}_{\mathrm{M}}$ and $\mathrm{FWHM}_{\mathrm{m}}$, the major and minor synthesised beam fitted FWHM
(column 9 in Table \ref{tab:observ}), respectively.}
\end{table*}

\subsection{The nature of SNe 2010cu and 2011hi}\label{subsec:11hi}

In \citet{erkki12} we investigated the nature of SNe 2010cu and 2011hi by means of five epochs of NIR photometric 
observations prior to 2012. Based on the comparison of the light curves of SN\,2010cu and SN\,2011hi to the CCSN templates 
presented in \citet{seppo01} and the well sampled NIR light curves of the canonical Type IIP SN\,1999em \citep{krisciunas09}, 
it was concluded that the best fit for both SNe was either a Type IIP or a IIn/L SN \citep[see][]{erkki12}. The ordinary, 
linearly declining template fit was previously excluded because such fit suggested an extremely early 
discovery and an unrealistically long rise time to the light curve peak. A NIR-bright Type IIn/L SN with $\sim$0\,mag 
of extinction was favoured in \citet{erkki12} as the type of SN\,2011hi due to the low ($<2$\,mag) total line-of-sight 
extinction implied by the \textit{J-K} colour map of the galaxy. However, a Type IIP light curve with higher extinction 
could also provide a good fit to the data, and as noted in \citet{erkki12}, higher localized extinctions are possible, as 
the colour map is tracing only the large scale structures of the LIRG.

\begin{figure*}
   \centering
\includegraphics[scale=0.5]{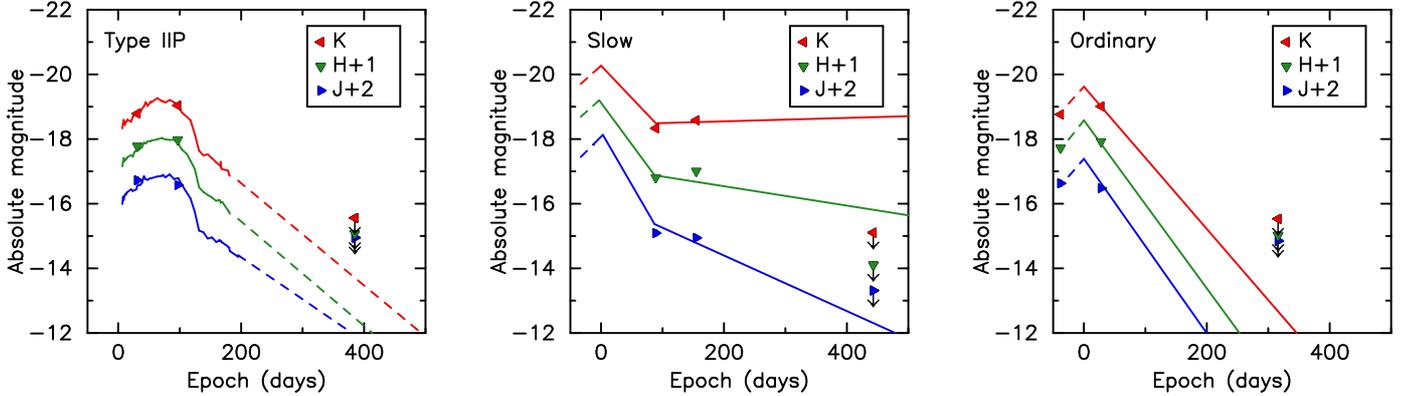}
   \caption{Template light curve fits for SN\,2011hi presented in \citet{erkki12} adopting the \citet{calzetti00} extinction
            law, with the addition of the late time \textit{JHK} upper limits obtained for SN\,2011hi in this work. The epochs 
            not covered by the templates are linearly extrapolated and marked with the dashed lines.}\label{fig:gemini}
\end{figure*}

In Figure \ref{fig:gemini} we present the template $\chi^{2}$ fits of the SN\,2011hi light curves adopting the 
\citet{calzetti00} reddening law and including the new upper limits. However, these limits are not included in the fit, 
i.e., the values reported in the Table 2 of \citet{erkki12} are unchanged. Based on the new data, the slowly declining 
(NIR-bright Type IIn/L) SN template is now excluded due to the fading of the SN. In addition to the SN\,1999em template, 
we have now also compared the NIR light curve of SN\,2011hi with that of other Type IIP SNe: 2003hn \citep{krisciunas09} 
and 2004et \citep{maguire10}. The results of the best fits are similar to those obtained with SN\,1999em, but having larger 
$\chi^{2}$ values. The \textit{V}-band extinctions from the three different Type IIP SNe used for comparison are consistent 
within 0.5\,mag for the adopted extinction law. We conclude that SN\,2011hi was in fact a slightly over-luminous (1--2\,mag 
brighter than SN\,1999em) Type IIP SN discovered roughly a month after the explosion, with 5 or 7\,mag of host galaxy 
extinction in the \textit{V}-band adopting either the \citet{cardelli89} or the \citet{calzetti00} reddening law, 
respectively.

There are no radio sources detected with the e-EVN above $3\sigma$ at/around the reported positions for either SN\,2010cu, 
nor SN\,2011hi. Therefore, the $3\sigma$ upper limits at 5\,GHz correspond to $2.4\times10^{27}$\,\lunits{} for both 
SNe. The SNe are also not detected in the archival 8.4\,GHz VLBA observations (project BC196) carried out two months 
after our e-EVN observations. Source B2a detected with e-MERLIN at 6.9\,GHz is positionally consistent within the uncertainties 
with SN\,2010cu. However, B2a is not a compact source, and we thus discard the possibility of it being a SN.

SN\,2010cu, has an estimated age of 14-17 months at the time of the radio observations made in
2011. Its non-detection at 5, 6.9 and 8.4\,GHz is an indication that: (i) the radio emission at 
these frequencies was well into the optically thin decline phase of its evolution (e.g., a Type Ib/c SN), 
now having a luminosity well below $2.4\times10^{27}$\,\lunits{}; and/or (ii) SN\,2010cu is an intrinsically 
low-luminosity Type II radio supernova (e.g., a Type IIP such as SN\,1999em) and thus more sensitive observations 
would be required for its detection.

SN\,2011hi was also not detected in our radio observations made 2-5 months after its explosion, and this 
is consistent with the fact that type IIP SNe are low-luminosity radio emitters. In fact, other 
well known type IIP SNe like SNe 1999em, 2004dj, and 2004et \citep[][respectively]{pooley02,beswick05,marti07} have 
peak radio luminosities below $10^{26}\,\lunits{}$ \citep[see e.g., figure 2 in][]{chevalier06}. At NIR wavelengths 
SN\,2011hi was only slightly brighter than SN\,1999em; if its radio emission was likewise only slightly stronger, we 
can expect that its radio luminosity was well below the 3$\sigma$ detection threshold ($2.4\times10^{27}$\,\lunits{}) 
of our most sensitive radio observations carried out with the e-EVN. We note that the non-detection of SNe 2010cu and 
2011hi at radio wavelengths does not imply that no radio emission has been produced by these NIR detected SNe.

To gain a better understanding of SNe in LIRGs, it is desirable to have a radio monitoring of the 
NIR-detected SNe including observations at different frequencies over many epochs. This strategy would 
help to detect either a fast or a slowly-evolving SN, which can also help to constrain their type \citep{stockdale07}.
However, not all CCSNe will be bright radio emitters. This depends on the interaction between the SN ejecta and the 
circumstellar medium, which gives rise to the synchrotron emission at radio frequencies \citep[see 
e.g.,][]{chevalier82}.

\subsection{The circumnuclear regions of IC\,883: a jet or a disk?}\label{sec:ic883_structure}

The quest to detect radio emission from SNe 2010cu and 2011hi has brought the opportunity to unveil different 
facets of IC\,883. In the following we discuss the circumnuclear structure of IC\,883 and the nature 
of its nuclear regions.

Our e-MERLIN observations reveal a double-sided structure with an approximate projected extent of 2\,arcsec 
($\sim1$\,kpc) by 0.3\,arcsec ($\approx150$\,pc), at a position angle of 130\degr{} (Figure \ref{fig:ic883zoom}). 
This inclination matches that previously found in other radio studies \citep[e.g.,][]{condon91,parra10}. However, the 
double-sided structure in IC\,883 is revealed for the first time at 6.9\,GHz with our e-MERLIN observations \citep[see also
the 1.4\,GHz MERLIN map from][]{clemens04}. VLA observations \citep[e.g., from][]{condon91,parra10} do not have enough spatial 
resolution to image the circumnuclear regions with similar level of detail. By convolving our e-MERLIN image with a beam of size 
0\farcs4, which corresponds to VLA A-configuration observations at 5\,GHz, we were able to recover a structure similar to that 
shown in figure 2 from \citet{parra10}.

At first glance, the structure seen with e-MERLIN resembles that of a two-sided jet of an AGN, whose restarting 
activity explains the presence of two sets of condensations (B1a-B1b and B2a-B2b; see left panel in Figure \ref{fig:ic883zoom}),
at either side of the core A. The larger flux density of the approaching components in the jet, B1a and B2a in this case, with 
respect to the flux density of the receding components, B1b and B2b, could be explained by Doppler boosting effects.

There are however strong arguments against a jet interpretation: (i) the approaching components are at shorter
distances from the core than the receding components, thus implying that their apparent velocities are smaller; (ii) the
putative jet structure is placed on top of the disk of the galaxy, and this is difficult to explain, unless we
consider that the directions of jet and disk coincide due to a projection effect.

Different studies of the molecular and atomic gas towards IC\,883 \citep{downes98,bryant99,clemens04}, give 
evidence of the rotation of the disk from South-East towards North-West. This situation would imply that the putative jet is 
perpendicular to the rotation axis, and hence no projection effect could explain the superposition of the jet on the disk. 
\citet{schmitt02} studied the orientation of jets with respect to the rotation axis in a sample of radio galaxies and 
found that large misalignments ($<77\degr{}$) may occur due to warping mechanisms in the black hole's accretion disk. In 
the case of IC\,883 the  misalignment would be close to 90\degr{}, so a jet scenario for the structure revealed with e-MERLIN
is even less viable.

If we consider instead a rotating disk (or ring) scenario, component A would be at the centre, 
while B1a, B2a (approaching components) and B1b, B2b (receding components) would be continuum radio components of 
the disk itself. We note that the line joining component B1a with B1b through A, is not the same as the line joining
B2a with B2b, which gives evidence of the disk/ring being warped \citep[in agreement with previous studies, e.g.,][]{clemens04},
probably due to the interaction with the ambient material. We note that the structure unveiled by e-MERLIN has a width of approximately 
150\,pc. Whilst such width might appear too narrow for representing a disk, thin disks of a few hundreds of parsecs 
are not uncommon for edge-on galaxies at radio frequencies \citep{dumke95}. Furthermore, e-MERLIN resolves out some 
of the extended emission which is otherwise visible at lower resolution, for instance with the VLA \citep{condon91, parra10}, 
and hence, the size of the structure mapped with e-MERLIN is compatible with a disk/ring structure.

Thus, whilst we cannot completely rule out the possibility that the structure mapped by e-MERLIN is a jet, 
there is strong evidence against it, and evidence favouring a rotating disk/ring nature instead.

Further e-MERLIN observations at different frequencies could be used to probe the radio lifetime of the 
condensations at either side of the core A. Observations optimized for polarimetry studies would yield information 
on the configuration of the large-scale magnetic field, which could probe the existence of shocked material as
expected in a jet scenario, hence helping to arrive to a firm conclusion.

\subsection{The ongoing nuclear activity in IC\,883}

Our e-EVN observations (see right panel in Figure \ref{fig:ic883zoom}), made simultaneous to our e-MERLIN observations,
reveal the presence of at least two compact sources (labelled as A1 and A2) above $5\sigma$ in the innermost 
nuclear regions of IC\,883. The high brightness temperature of sources A1 and A2 indicates a non-thermal origin for both 
of them.

The brightest component in our e-MERLIN image (A) has a peak intensity similar to that of A1. By adding up the 
flux densities of components A1 and A2 in our e-EVN map, we recover a flux density of $4.38\pm0.27$\,mJy, which is still 
lower than both the peak intensity and the flux density of component A (see Table \ref{tab:ic883_radsou}). We find that
A1 has a strong contribution to the flux density at larger scales (although seen at slightly different frequencies), and 
furthermore, has a position highly coincident with component A, despite the use of different phase reference sources and 
different spatial resolutions. Hence, A1 is very likely the main powering source for the radio emission of component A.

From our e-EVN observations alone we cannot characterise the nature of sources A1 and A2, nor investigate a 
possible relation between components A1 and A. We have therefore searched for additional, publicly available, archival 
VLBI data (see Table \ref{tab:observ}) that could be of use to investigate these issues.

Component A1 from our e-EVN observations is detected in all the additional VLBI observations (see Figure 
\ref{fig:allvlbi}) and appears to fluctuate in flux density at 5\,GHz, varying for at least a 
factor of three between the minimum and maximum value displayed within the seven years comprised by the different VLBI 
epochs (see Table \ref{tab:ic883_radsou}). Component A2 is not detected in any other epoch. At 8.4\,GHz (experiment BC196) 
we detect a 5$\sigma$ source (A3) North-West of component A1. The resolution of our e-EVN observations (RR006) does not 
allow us to detect component A3, but the source is also not detected in any of the other VLBA epochs at 5\,GHz, which have a 
resolution similar to that of experiment BC196. A 3$\sigma$ component placed North-East from component A1, is detected as a 
5$\sigma$ source in experiment EP055 and is labelled as A4. This source has no counterpart at 8.4\,GHz in the observations
under project BC196, indicating that at that frequency the emission from A4 has become optically thin, as expected in a SN 
scenario. We thus have a total of three transient sources (A2, A3 and A4) in the nucleus of IC\,883, in addition to the 
more luminous, variable, component A1, which has remained visible for at least seven years.

\subsection{The nature of the compact sources in the IC\,883 nucleus}\label{subsec:nat_cc}

\begin{figure}[ht]
   \centering
\includegraphics[bb=90 238 500 587,clip,scale=0.62]{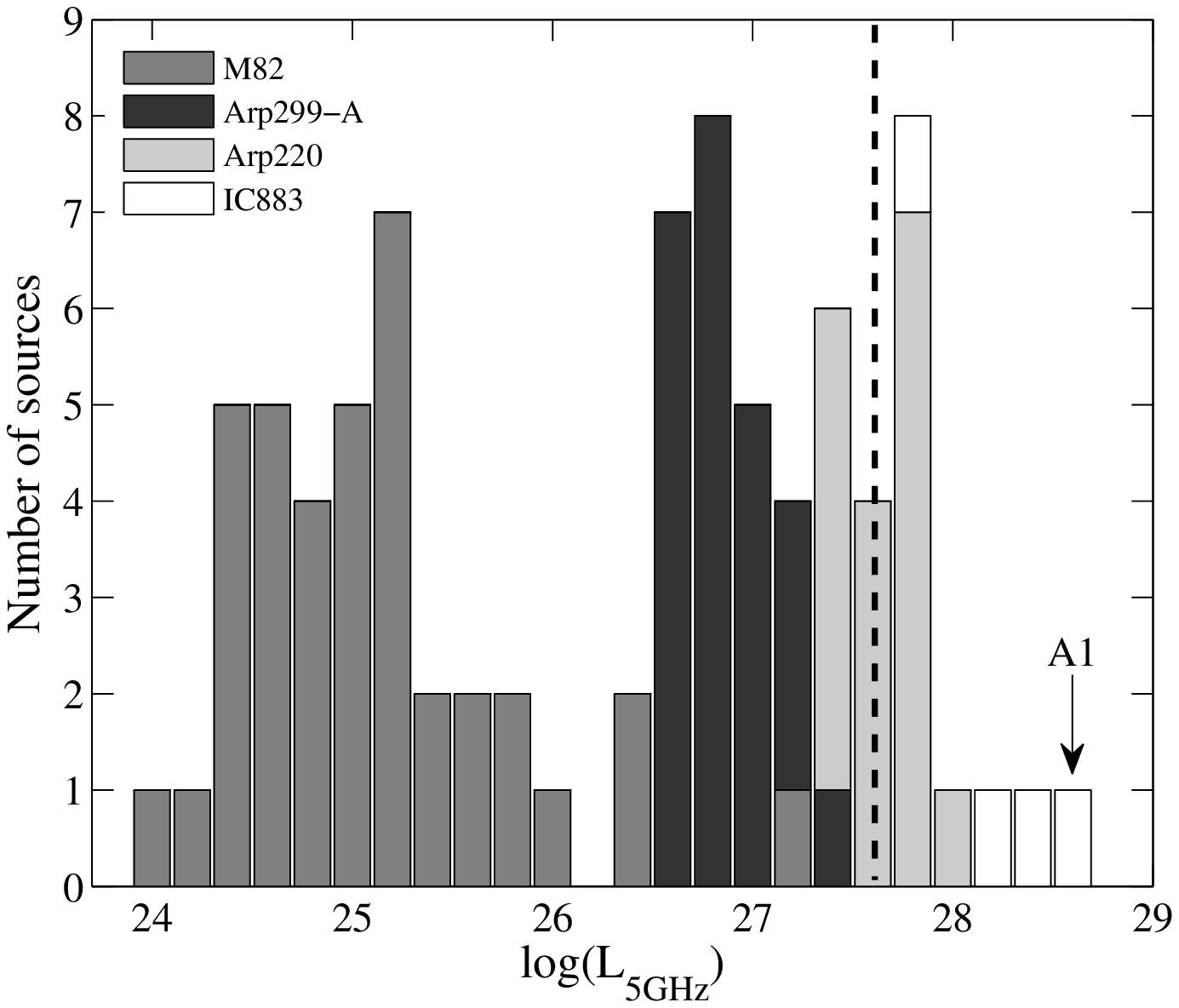}
   \caption{Luminosity distribution at 5\,GHz of SNe and SNRs in M82, Arp\,299-A, Arp\,220. All the IC\,883 nuclear sources have 
been added for comparison purposes. Each bin has a width of log$_{10}(L_{5\,\mathrm{GHz}})=0.2$ with $L_{5\,\mathrm{GHz}}$ in 
units of \lunits{}. The SN and SNR luminosities have been retrieved from \cite{fenech08} and \citet{sn2008iz} for M82, 
\citet{bondi12} for Arp\,299-A and from \citet{batejat11} for Arp\,220. The most luminous M82 source corresponds to SN\,2008iz. 
In the case of Arp\,220 we have included SNe and SNRs from both West and East nuclei. The luminosities of the IC\,883 components 
were taken from Table \ref{tab:ic883_radsou}: from experiment RR006 for A1 and A2, from experiment BC196 for A3, and from experiment 
EP055 for A4. We have assumed a spectral index of $-0.7$ to obtain the luminosity at 5\,GHz of component A3. Note that A1 is 
the brightest source, and with our most sensitive VLBI image (RR006) we are only sensitive to sources above a 5$\sigma$
luminosity limit of $4\times10^{27}$\,\lunits{}, indicated by a dashed line in the histogram.}\label{fig:histo}
\end{figure}

The milliarcsec sources detected in the innermost regions of IC\,883 give evidence of its ongoing nuclear activity.
A variable source, A1, has been detected in six different VLBI epochs
(see Table \ref{tab:ic883_radsou} and Figure \ref{fig:allvlbi}) throughout seven years, and appears to be the main 
powering source of the emission at larger scales imaged with e-MERLIN. Three transient sources have also been detected. 
All the sources have sizes and brightness temperatures indicating a non-thermal origin. The transient sources are most 
probably SNe and/or SN remnants (SNR), whereas for component A1 the nature is less clear. To investigate the nature of 
sources A1--A4, we compare their luminosities with those reported in the literature for the non-thermal sources (SNe 
and SNRs) at 5\,GHz in the starburst galaxy M82, the LIRG Arp\,299-A and in the ULIRG Arp\,220 (Figure \ref{fig:histo}).

From Figure \ref{fig:histo} we see that the nuclear components of IC\,883, although being less numerous, 
are in general more luminous than the components identified as SNe and SNRs in the nuclear regions of other galaxies 
with high star formation rates. The lifetime of the transient sources in IC\,883 ranges from less than a few months
up to a few years. It is thus very likely that these sources are a mixture of SNe and SNRs, therefore 
indicating a recent starburst activity in the nucleus of IC\,883. We lack spectral information that could have helped 
to differentiate between a SN or a SNR origin for each of these transient sources.

Component A1 is the most luminous of all the sources (see histogram in Figure \ref{fig:histo}). A SN or a SNR 
origin are not very likely, based on its flux density fluctuations (see columns 5 and 6 in Table \ref{tab:ic883_radsou}), 
its longevity, and its relatively flat two-point spectral index between 5 and 8.4\,GHz, 
$\alpha = 0.3$ ($S_{\nu} \sim \nu^{\alpha}$), calculated from the flux densities obtained in experiments RR006 
and BC196. Therefore A1 is very likely an AGN.

Further evidence for an obscured AGN in IC\,883 can be inferred from the ratio of radio to hard 
X-ray luminosity \citep[$R_{\mathrm{X}}=\nu L_{\nu}(5\,\mathrm{GHz})/L_{\mathrm{HX}}$;][]{terashima03}. \citet{iwasawa11} 
report a hard X-ray luminosity ($L_{\mathrm{HX}}=L[2$--$10\,\mathrm{keV}]$) of $6.4\times10^{40}$\,\rlunits{} for IC\,883. 
Considering the luminosity at 5\,GHz of A1 observed with the e-EVN (see Table \ref{tab:ic883_radsou}), we calculate 
$R_{\mathrm{X}}\approx3.6\times10^{-3}$, which is consistent with a low-luminosity AGN (LLAGN) or a normal AGN in 
a radio galaxy, as inferred from figure 4 of \citet{terashima03}.

By comparing the maximum variability (a factor of $\sim$3) displayed by A1 in our observations, with the largest 
inter-year variability displayed by typical LLAGNs \citep{llagnvar}, we find that A1 would be placed at the high end of 
the LLAGN behaviour. Thus, we can only conclude that A1 is a candidate for either a normal AGN or a low-luminosity one.

Component A4 whose elongation could indicate a sort of outflow, has however no counterpart 
at 8.4\,GHz (experiment BC196). Future VLBI observations at different frequencies should allow an unequivocal
characterisation of the AGN candidate in IC\,883, as well as possibly discovering further nuclear components.

If we consider sources A2, A3 and A4 as having their origin in recent star formation activity, we would have
three CCSN events that have occurred in the period between June 2006 and May 2011, i.e., five years. We can thus estimate 
a CCSN rate for the IC\,883 nuclear region of $\approx 0.6_{-0.3}^{+0.6}$\,yr$^{-1}$, using the upper and lower Poisson 
$1\sigma$ uncertainties for three events given by \citet{gehrels}. Combining the nuclear CCSN rate with the CCSN rate 
inferred from the detection of SNe 2010cu and 2011hi within 3.8\,yr in the circumnuclear regions of IC\,883, we obtain a 
CCSN rate for the entire galaxy of $\approx 1.1_{-0.6}^{+1.3}$\,yr$^{-1}$. These rates are considered as lower limits 
since our irregular temporal sampling both at radio and at NIR wavelengths might have missed rapidly evolving SNe, and also 
because our 5$\sigma$ radio luminosity detection threshold ($4\times10^{27}$\,\lunits{}) prevents us from detecting Type IIP 
and IIb SNe, and even some Type IIL. We note that the total rate estimate is in agreement with the IR luminosity-based CCSN 
rate estimate (see Section \ref{sec:intro}). It is thus of great interest to estimate the contribution of a putative AGN to 
the IR luminosity in order to obtain a more reliable estimate of the expected CCSN rate.

\subsection{The infrared SED of IC\,883: estimating the contribution of an AGN}

As mentioned in Section \ref{sec:intro}, IC\,883 has been classified as a composite starburst-AGN, 
based on optical diagnostic diagrams \citep{yuan10}.

\begin{figure}[ht]
   \centering
\includegraphics[bb=67 406 520 770,clip,scale=0.5]{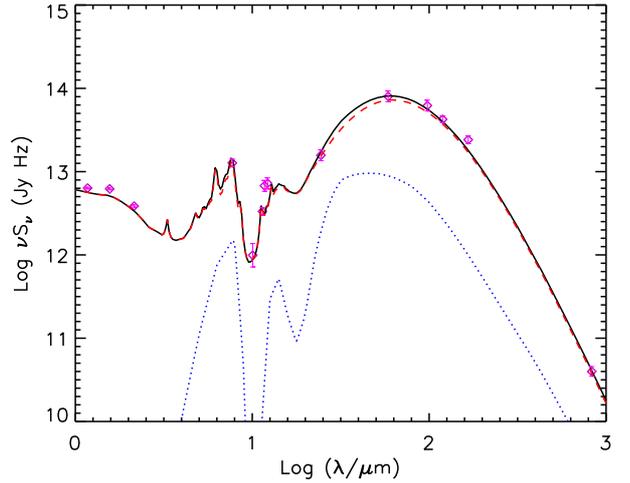}
   \caption{The infrared SED of IC\,883 including 2MASS and {\it IRAS}
   data, as well as the SCUBA point from \citet{dunne00}. The red-dashed
   line is our starburst model fit, and the blue-dotted line is the model fit
   for an AGN upper limit contribution. 
   The black-solid line is the model fit including both components.}\label{fig:ic883sed}
\end{figure}

We have modelled the spectral energy distribution (SED) of IC\,883 (see Figure \ref{fig:ic883sed}) 
to estimate the contribution of an AGN to the total IR luminosity and to provide an estimate for the CCSN rate
in this galaxy. For this we used a grid of AGN dusty torus models that have been computed with the method of 
\citet{efs95} and a grid of starburst models that have been computed with the method of \citet{efs00},
but with a revised dust model \citep{efs09}. The spatial resolution of the mid- and far-IR data used for 
the modelling is not high enough to spatially separate the different components. Therefore to distinguish 
the contribution of the AGN torus and the starburst we have fitted a model combining both an AGN and a 
starburst.

For the AGN, we considered a grid of tapered disc models computed with the method of 
\citet{efs95}. In this grid of models we consider four discrete values for the equatorial 1000\,\r{A} optical 
depth (500, 750, 1000, 1250), three values for the ratio of outer to inner disc radii (20, 60, 100) and three 
values for the opening angle of the disc (30\degr{}, 45\degr{} and 60\degr{}). The best fit parameters which
resulted in the minimum $\chi^{2}$ when we combine the starburst and the AGN torus models, are an opening angle of 
60\degr{} for the torus, an equatorial optical depth of 1250 at 1000\,\r{A}, and a ratio of outer to inner
radius of 60. The AGN torus is viewed close to edge-on due to the IC\,883 geometry; thus, its apparent 
contribution to the total infrared emission is insignificant (probably $<$10 per cent of the starburst 
luminosity). Therefore the starburst clearly dominates the observed luminosity of the entire galaxy. We note 
that due to viewing the torus almost edge-on and due to its extremely high optical depth, the intrinsic luminosity 
of the AGN itself could be a factor of $\sim$30 times higher than the observed one. This enhancing factor is
related to the anisotropy of the torus emission \citep[see][for a definition]{efs06}, and implies that the intrinsic 
luminosity of the AGN could be about three times higher than the observed starburst luminosity.

Our model incorporates the stellar population synthesis model of \citet{bruzual03}, which makes a 
prediction of the SN rate, $\snrate(t)$,  at a time $t$ after star formation in an instantaneous burst. 
The starburst model of \citet{efs00} predicts the spectrum of this instantaneous burst at time $t$ and assumes 
a star formation history for the starburst. We assume that the SFR in IC\,883 declines exponentially with an 
e-folding time of 20\,Myr, and that the age of the starburst is 55\,Myr. The initial optical depth of the 
molecular clouds that constitute the starburst is assumed to be 100. The SFR averaged over the duration of the 
starburst is 185\,M$_{\odot}$\,yr$^{-1}$. It is also possible to calculate self-consistently SN rate at different 
stages in the evolution of a starburst by convolving the star formation history with $\snrate(t)$. This yields a 
$\snrate$ of 1.1\,yr$^{-1}$ which is relevant for the entire galaxy and can therefore be compared with the observed 
values. The combination of the CCSN rate for the nuclear regions based on three radio SNe, and the CCSN rate for the
circumnuclear regions based on the discovery of SNe 2010cu and 2011hi, resulted in a lower limit for the CCSN rate of 
$\approx 1.1_{-0.6}^{+1.3}$\,yr$^{-1}$ for the entire galaxy (see Section \ref{subsec:nat_cc}). This CCSN rate estimate 
is in good agreement with the predicted CCSN rate for the entire IC\,883 galaxy from the IR SED modelling.

IC\,883 being less luminous in the IR, but farther away than Arp\,299-A and in a much more advanced merger stage, 
contains both a starburst and an AGN. The nuclear starburst in IC\,883 is apparently less prolific than in Arp\,299-A 
\citep{paper1}. However, the AGN in IC\,883 seems to be stronger than the one in Arp\,299-A \citep{paper2}, and even 
powering the radio luminosity at circumnuclear scales, although its contribution to the emission of the entire galaxy 
at IR wavelengths is apparently insignificant.

\section{Summary}\label{sec:sum}

We have imaged the nuclear and circumnuclear regions of the LIRG IC\,883 at radio frequencies. Our 
e-EVN observations, together with archival VLBI observations, reveal the presence of at least four non-thermal 
components in the nuclear regions of IC\,883. Three of these are transient sources and one is a long-lived, 
variable compact source, which is very likely an AGN. Our e-MERLIN observations highlight the presence of a striking 
double-sided structure, likely representing a warped disk/ring, in agreement with previous studies on the interstellar 
medium of this galaxy.

The source we identify as an AGN candidate is powering the radio emission in IC\,883 at nuclear
and circumnuclear scales, and yet, as indicated by our radio and NIR observations, as well as by our model of 
the IR SED, this LIRG displays very active star formation both within the innermost 100\,pc of its nucleus, and 
in the circumnuclear regions. We have estimated lower limits for the CCSN rate in the innermost nuclear regions of 
$0.6_{-0.3}^{+0.6}$\,yr$^{-1}$ based on the detection of three radio SNe above a 5$\sigma$ luminosity threshold of 
$4\times10^{27}$\,\lunits{}, and $1.1_{-0.6}^{+1.3}$\,yr$^{-1}$ for the entire galaxy, based on the nuclear 
radio SNe and the NIR SNe 2010cu and 2011hi discovered in the circumnuclear regions. Our most recent NIR 
Gemini-North observations of IC\,883 clarify the nature of SN\,2011hi as a Type IIP SN, in agreement with
its non-detection in our observations at radio frequencies.

The composite starburst-AGN nature of IC\,883, revealed previously by optical spectroscopy 
\citep[e.g.,][]{yuan10}, is now supported by our radio observations for the first time, thus making IC\,883
one of the few galaxies where the starburst-AGN connection can be studied in detail.

\begin{acknowledgements}
We thank the anonymous referee for his/her comments. The authors are thankful to the EVN Directors for their rapid 
response for ToO observations. The European VLBI Network is a joint facility of European, Chinese, South African 
and other radio astronomy institutes funded by their national research councils. We also made use of observations 
with e-MERLIN, the UK's facility for high resolution radio astronomy observations, operated by The University of Manchester 
for the Science and Technology Facilities Council; and observations obtained at the Gemini Observatory, 
which is operated by the Association of Universities for Research in Astronomy, Inc., under a cooperative 
agreement with the NSF on behalf of the Gemini partnership: the National Science Foundation (United States), 
the Science and Technology Facilities Council (United Kingdom), the National Research Council (Canada), CONICYT 
(Chile), the Australian Research Council (Australia), Minist\'{e}rio da Ci\^{e}ncia e Tecnologia (Brazil) and Ministerio 
de Ciencia, Tecnolog\'{i}a e Innovaci\'{o}n Productiva (Argentina). This article is also based on observations 
made with the Very Long Baseline Array (VLBA) of the National Radio Astronomy Observatory (NRAO); the NRAO is a 
facility of the National Science Foundation operated under cooperative agreement by Associated Universities, Inc. 
This work made use of the Swinburne University of Technology software correlator, developed as part of the Australian 
Major National Research Facilities Programme and operated under licence. Our work is supported by the European Community 
Framework Programme 7, Advanced Radio Astronomy in Europe, grant agreement no.: 227290. C.R.-C., M.A.P.-T. and A.A. 
acknowledge financial support from the Spanish MICINN through grant AYA2009-13036-C02-01, co-funded with FEDER funds. 
S.M., E.K. and C.R.-C. acknowledge financial support from the Academy of Finland (project: 8120503). E.K. acknowledges 
support from the Finnish Academy of Science and Letters (Vilho, Yrj\"{o} and Kalle V\"{a}is\"{a}l\"{a} Foundation).
\end{acknowledgements}


\end{document}